\PassOptionsToPackage{unicode}{hyperref}
\PassOptionsToPackage{hyphens}{url}
\documentclass[]{article}
\usepackage[margin=1in]{geometry}
\usepackage{xcolor}
\usepackage{graphicx}
\usepackage{float}
\usepackage{caption}
\usepackage{amsmath,amssymb}
\usepackage{iftex}
\ifPDFTeX
  \usepackage[T1]{fontenc}
  \usepackage[utf8]{inputenc}
  \usepackage{textcomp}
\else
  \usepackage{unicode-math}
  \defaultfontfeatures{Scale=MatchLowercase}
  \defaultfontfeatures[\rmfamily]{Ligatures=TeX,Scale=1}
\fi
\usepackage{lmodern}
\IfFileExists{upquote.sty}{\usepackage{upquote}}{}
\IfFileExists{microtype.sty}{%
  \usepackage[]{microtype}
  \UseMicrotypeSet[protrusion]{basicmath}
}{}
\usepackage{fancyvrb}
\usepackage{fvextra}
\DefineVerbatimEnvironment{Highlighting}{Verbatim}{fontsize=\small,breaklines,breakanywhere}
\newenvironment{Shaded}{}{}
\usepackage{longtable,booktabs,array}
\usepackage{calc}
\usepackage{etoolbox}
\makeatletter
\patchcmd\longtable{\par}{\if@noskipsec\mbox{}\fi\par}{}{}
\makeatother
\IfFileExists{footnotehyper.sty}{\usepackage{footnotehyper}}{\usepackage{footnote}}
\makesavenoteenv{longtable}
\providecommand{\tightlist}{%
  \setlength{\itemsep}{0pt}\setlength{\parskip}{0pt}}
\usepackage{bookmark}
\IfFileExists{xurl.sty}{\usepackage{xurl}}{}
\hypersetup{
  hidelinks,
  pdfcreator={LaTeX via build_paper.py}}
\title{The Single-File Test: A Longitudinal Public-Interface Evaluation of First-Output LLM Web Generation with Social Reach Tracking}
\author{Diego Cabezas Palacios\\\texttt{diegocp01@gmail.com}}
\date{}

\begin{document}

\maketitle

\begin{center}
\textbf{Code and materials:} \url{https://github.com/diegocp01/html\_ai\_battle}
\end{center}

\begin{abstract}

Standard coding benchmarks often reward local correctness while missing the broader challenge of producing a complete, \textbf{visually coherent}, interactive web application on the \textbf{first try}. This paper presents an \textbf{eight-week longitudinal} observational comparison of \textbf{68 single-file HTML generations} collected across \textbf{17 public experiments} in the "\textbf{HTML AI Battle}" project between \textbf{December 10, 2025 and February 4, 2026}. The study compares four reasoning model families, \textbf{GPT, Gemini, Grok, and Claude}, under a strict fixed user protocol using \textbf{public chat interfaces} with \textbf{no custom instructions}, no personality tuning, and \textbf{no repair prompts}. Each output is \textbf{evaluated} with both \textbf{human scores} and an \textbf{LLM-as-a-judge} layer for prompt adherence, functional correctness, and UI quality, then packaged into a \textbf{shared social-media protocol} spanning X (Twitter), TikTok, and YouTube. Using the results tracker, the paper also fits \textbf{two supervised predictive analyses}: an experiment-level model that tests whether pre-publication technical and audio variables \textbf{predict 24-hour X (Twitter) impressions}, and a generation-level model that tests whether \textbf{prompt wording} and \textbf{model family} predict \textbf{HTML verbosity}.

The main results are straightforward. Under this \textbf{fixed user protocol}, \textbf{Claude was the strongest and most consistent family}, leading mean performance and winning \texttt{9/17} prompts under the primary human weighted score. Across all model families, \textbf{longer measured reasoning time was not associated with higher quality} overall. \textbf{Gemini as a judge} was significantly \textbf{more lenient than the human evaluator} on functional correctness and overall performance, but with only \texttt{17} experiment-level comparisons the study was underpowered to rule in or rule out stable self-favoring bias. The exploratory \textbf{X-impressions model} remained weak under post-screen cross-validation (\texttt{MAE\ =\ 46,874}, \texttt{R\^{}2\ =\ -0.377}), and with only \texttt{17} experiments those estimates should be read as unstable descriptive evidence of limited predictive signal rather than as a precise forecasting result. By contrast, the \textbf{HTML-lines model} performed materially better, with a simple model-family-only baseline outperforming prompt-aware alternatives (\texttt{MAE\ =\ 135.2}, \texttt{R\^{}2\ =\ 0.576}). Taken together, the findings suggest that the selected pre-publication technical/audio variables were \textbf{not sufficient to predict 24-hour X (Twitter) reach} in this dataset, while \textbf{code verbosity} was driven much more by \textbf{model family} than by prompt wording alone. These comparisons remain observational and are limited by public-interface drift, access-path differences, and one primary human scorer.

\end{abstract}

\section{Introduction}\label{introduction}

\textbf{Large language models can now generate working code in a single pass}, but many established coding evaluations focus on whether code is technically correct rather than whether the resulting application is usable, visually coherent, and interactive for a real user [1, 3, 4]. Web-oriented benchmarks have also emphasized that realistic browser tasks and visually grounded interfaces expose failures that text-only or local-code checks can miss [10, 11]. A browser-based toy, game, or simulation can fail in ways that do not appear in unit tests: broken controls, weak motion, confusing layouts, or visually incoherent interfaces. For end users, especially \textbf{non-technical users}, those failures matter as much as syntax.

The project was motivated by a \textbf{common pattern in public AI demos}: side-by-side comparisons often show several model outputs for the same prompt, but omit operational details such as the exact model variant, reasoning mode, latency, prompt length, generated code length, scoring rubric, and posting context. Those missing details make the comparisons entertaining but difficult to audit or analyze. HTML AI Battle was designed as a more structured version of that format: the same public-facing prompt is sent to several model families, the first output is preserved, and the surrounding metadata are tracked so that the final analysis can ask not only which output looked best, but also which \textbf{measured variables} were associated with quality, verbosity, and social reach.

This paper studies that gap through a longitudinal, \textbf{first-output-only} observational comparison of \textbf{public chat interfaces} focused on \textbf{single-file web applications}. Across 17 experiments, the \textbf{same task prompt} was given to \textbf{multiple frontier model families}, and the first delivered HTML file was treated as final. The study was intentionally run through public chat interfaces rather than APIs, with the exact access paths and interface constraints documented in the experimental protocol. The goal is therefore not a clean causal ranking of intrinsic model capability, but a fixed-protocol comparison of what these \textbf{public interfaces} delivered to a real user over time.

The data collection workflow was supported by the repository's local Data Collection Program [5]. That tool helped standardize repetitive steps such as experiment setup, timing, reasoning-word counts, HTML-line counts, scoring prompts, song-prompt generation, and post-template preparation. For each experiment $e$, the workflow produced \textbf{four model-level HTML artifacts}, one per model family. Each artifact was rendered in a browser, recorded as a video, and \textbf{scored from the recording} rather than from the source code. The social layer preserved the same paired structure by arranging the four rendered recordings into a \texttt{2\ x\ 2} comparison matrix. Each quadrant was labeled with the specific model variant shown in that cell, rather than with only a broad family name; examples include \texttt{GPT-5.2\ Extended\ Thinking}, \texttt{Gemini\ 3\ Pro}, \texttt{Grok\ 4.1\ Thinking}, and the LMArena label \texttt{Opus\ 4.5\ Thinking\ 32K}. The composite video was then paired with a single AI-generated song whose lyrics encoded the human observations and scores. The lyric-generation LLM was tracked and varied across experiments: \texttt{gemini-3-pro} was used for \texttt{6} songs, \texttt{opus-4.5-thinking-32k} for \texttt{6}, and \texttt{gpt-5.2-extended-thinking} for \texttt{5}. In each case, the song prompt asked the lyric model to use phonetic or nickname-style model names so the comparison would remain understandable in audio form. The final composite was posted on a twice-weekly Wednesday/Saturday schedule. This design keeps the unit structure explicit: four model-level outputs form one experiment-level social post.

The work addresses three research questions. First, \textbf{how do model families differ} in quality, latency, code verbosity, and reasoning efficiency when forced to produce a single-file application with no iteration? Second, when Gemini is used as a blind secondary judge, \textbf{does it behave as a neutral evaluator} or show evidence of systematic self-favoring, a concern related to broader LLM-as-a-judge evaluation work [2]? Third, what can be predicted from the tracker using supervised machine-learning models: can pre-publication technical and audio variables \textbf{predict 24-hour X (Twitter) impressions},\footnote{The platform is referred to as X in the dataset and analysis, with Twitter included parenthetically for reader recognition.} and can \textbf{prompt wording plus model family predict HTML output length?}

The main contributions are:

\begin{enumerate}
\tightlist
\item A fixed-protocol observational comparison of public chat interfaces on one-shot single-file web generation under real user conditions.
\item Evidence that, within this protocol, Claude was the strongest and most consistent family while longer reasoning did not improve quality overall.
\item A judge-comparison analysis showing that Gemini as a video-based secondary judge was more lenient than the human evaluator, while stable self-favoring bias remained unresolved in this small sample.
\item Two supervised predictive analyses using the same tracker: an experiment-level X-impressions model and a generation-level HTML-lines model, showing that technical metrics did not predict 24-hour X reach well in this dataset, while model family did predict HTML verbosity.
\end{enumerate}

\section{Preliminaries}\label{preliminaries}

\subsection{Scoring Framework}\label{scoring-framework}

Throughout the paper, $i$ indexes model-level generations, while $e$ indexes experiment-level aggregates. In this dataset, each experiment $e$ contains four generations $i$, one for each model family.

Each generated application was scored along the same three evaluation dimensions:

\begin{itemize}
\tightlist
\item Prompt Adherence (\texttt{PA}): how completely the output satisfied the requested constraints.
\item Functional Correctness (\texttt{FC}): whether the application actually worked as intended.
\item UI Quality (\texttt{UI}): visual clarity, layout, polish, and readability.
\end{itemize}

For generation $i$, the score vector is

\[
\mathbf{s}_i =
\begin{bmatrix}
PA_i \\
FC_i \\
UI_i
\end{bmatrix}
\in \mathbb{R}^3.
\]

This representation supports geometric comparisons between evaluations. Given two score vectors, directional agreement can be measured with cosine similarity:

\[
\cos(\theta_{ab}) =
\frac{\mathbf{s}_a^\top \mathbf{s}_b}
{\|\mathbf{s}_a\| \, \|\mathbf{s}_b\|}.
\]

As a diagnostic example, the first two model-level score vectors in the notebook had a cosine similarity of \texttt{0.9966}, indicating nearly identical score-profile direction for that pair. This pairwise value is not treated as a central result because it reflects only one comparison; the main score-structure result below uses all \texttt{68} model-level outputs.

Because the three score dimensions are not independent in practice, I also model their shared structure through the covariance matrix

\[
C = \frac{1}{n-1} Q^\top Q,
\]

where $Q$ is the centered $n \times 3$ matrix of score vectors. The eigendecomposition

\[
C \mathbf{v}_k = \lambda_k \mathbf{v}_k
\]

reveals whether a dominant co-movement pattern exists. In the model-level score matrix from this study, the first eigenvalue explained \texttt{92.8\%} of the total variance across the \texttt{68} outputs, showing that prompt adherence, functionality, and UI quality co-move strongly in this dataset. That pattern is consistent with a shared overall quality dimension, but it could also reflect rubric overlap or halo effects from a single human scorer, so it should be read as descriptive rather than as proof of a single latent construct.

\subsection{Experimental Protocol}\label{experimental-protocol}

The study uses a strict \textbf{first-output-only} protocol designed to approximate a realistic end-user interaction. Each experiment begins with one natural-language prompt requesting a complete, \textbf{single-file HTML/CSS/JavaScript application}. Across the \texttt{17} experiments, prompt length ranged from \texttt{10} words to \texttt{123} words, with a median of \texttt{20} words. The same prompt was issued once to each model family, with \textbf{no retries}, debugging prompts, repair prompts, or iterative steering; the \textbf{first delivered output was treated as final}. The raw generated HTML was preserved without human code edits before execution and scoring. The intended collection sequence was GPT, Gemini, Grok, and then Claude through LMArena; tracker timestamps follow that nondecreasing order in \texttt{14/17} experiments, with a few early exceptions or same-minute ties, so collection order is documented as workflow context rather than used as an analytical variable.

The dataset contains \texttt{68} generations across \texttt{17} experiments. The comparison is organized at the family level:

\begin{itemize}
\tightlist
\item \texttt{GPT}, collected at \texttt{chatgpt.com} and recorded in the tracker as \texttt{gpt-5.1-extended-thinking} or \texttt{gpt-5.2-extended-thinking} as the public interface evolved. The public interface labels were GPT-5.1 or GPT-5.2, with Thinking and extended reasoning effort selected.
\item \texttt{Gemini}, collected at \texttt{gemini.google.com} and recorded in the tracker as \texttt{gemini-3-pro}: the public interface label was Gemini 3, with Pro selected from the model selector.
\item \texttt{Grok}, collected at \texttt{grok.com} and recorded in the tracker as \texttt{grok-4.1-thinking}: the public interface label was Grok 4.1, with Thinking selected from the model selector.
\item \texttt{Claude}, collected through \texttt{lmarena.ai} with exact LMArena label \texttt{opus-4.5-thinking-32k}.
\end{itemize}

The GPT, Gemini, and Grok entries above are tracker labels rather than exact product names; their suffixes record the model-selector or effort settings used during collection. The Claude entry preserves the exact LMArena label used during collection.

All runs were collected through public user-facing interfaces rather than API orchestration. GPT, Gemini, and Grok were used with default settings, \textbf{no custom instructions}, and no personality tuning. Claude-family runs used Claude Opus through LMArena\footnote{The platform used during data collection was called LMArena; it is now available as arena.ai. LMArena is a community-driven model-comparison platform associated with Chatbot Arena-style evaluation: users can compare model outputs, vote on preferred responses, and contribute to public leaderboards [2]. This study did not use LMArena's anonymous battle mode for scoring. It used the platform's direct-chat access path to run the Claude Opus model variant recorded in the tracker (\texttt{opus-4.5-thinking-32k}). The paper therefore uses \texttt{Claude} as the model family label and \texttt{Claude\ Opus} as the specific model variant/access path.} instead of Anthropic's native interface. The repository's Data Collection Program [5] was used throughout this workflow to record the prompt, response timing, reasoning timing, generated HTML, rendered-output observations, and scores, and to provide templates for LLM-as-a-judge scoring prompts, song-lyrics prompts, and posting text. The 24-hour platform metrics were then recorded manually in the tracker after the fixed collection window. Although the user workflow, access websites, posting schedule, and 24-hour collection window were \textbf{standardized}, the study remains observational because public model availability, provider-side interface behavior, and platform context could not be experimentally randomized or held constant across the full collection period.

The resulting tracker contains \texttt{48} columns spanning the full workflow. At the run level, it records the experiment identifier, model label, model family, platform/access context, prompt, experiment type, visual type, generation date and time, input word count, and whether code appeared in the visible reasoning. For the model interaction and artifact, it records visible reasoning time, total response time, reasoning word/character/sentence counts, the top repeated reasoning keyword, generated HTML line count, rendered-output observations, and the human and Gemini sub-scores for prompt adherence, functional correctness, and UI quality, from which the tracker computes human and Gemini weighted performance scores. For social packaging, it records song voice gender, key, BPM, style, lyric word count, lyric-generation model, Suno version, posting date, posting day of week, X posting time, and video duration. For public outcomes, it records follower counts at posting time and 24-hour X impressions, likes, and shares, alongside TikTok views and likes and YouTube Shorts views and likes.

\begin{figure}[H]
\centering
\includegraphics[width=0.55\linewidth]{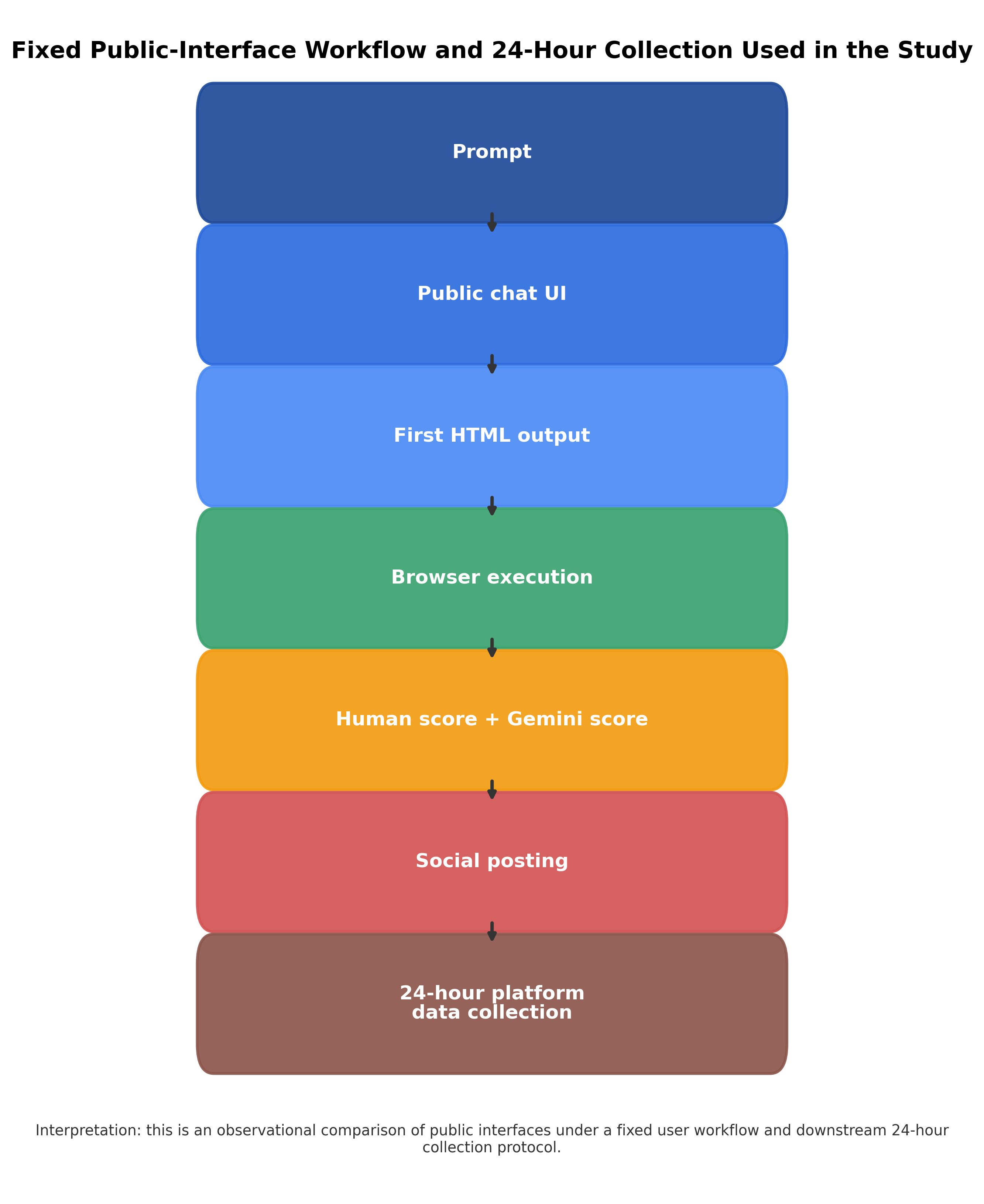}
\caption{Protocol overview of the fixed public-interface workflow and subsequent 24-hour platform-metrics collection step used throughout the study. The repository's Data Collection Program recorded workflow data such as prompt, timing, output, scoring, and posting-template fields; 24-hour platform metrics were manually entered in the tracker after the fixed collection window.}
\label{fig:figure-28}
\end{figure}

After generation, \textbf{each HTML file was executed locally in a standard browser} environment and \textbf{recorded as video}. Both the human evaluator and Gemini \textbf{judged the rendered browser recording only}, not the underlying HTML/CSS/JavaScript code. In other words, \textbf{scoring was based on videos} of each generated app running in the browser. The author served as the human evaluator and assigned the primary human scores using the three rubric dimensions above. The author was \textbf{not consistently blinded to model identity} during human scoring: in some cases the model identity was known, while in others it was not. These scores should therefore be interpreted as primary author judgments rather than fully blinded independent annotations. \textbf{Gemini was used as a secondary judge} because, at the time of evaluation, it was the available model in this workflow that \textbf{could accept video input}.\footnote{The judge comparison is therefore a video-based evaluation of rendered browser behavior, not a source-code review.} Gemini scored the videos (HTML outputs) in \textbf{four separate blind chats per experiment} using the same Data Collection Program-generated judging prompt template,\footnote{The Gemini judging prompt template was generated by the repository's Data Collection Program and reused across evaluations, with the experiment prompt and video supplied for the specific run. The template is reproduced in Appendix A.4.} the video evidence, and the original task prompt; \textbf{Gemini was not told which model produced each application}. Human sub-scores used a \texttt{0-10} scale with decimal values allowed, while Gemini was prompted for \texttt{1-10} integer scores. Because the human evaluator was only partially blinded while Gemini was blinded, and because the two scoring scales differed, Gemini-human comparisons should be interpreted as directional agreement and leniency checks rather than as a perfectly controlled judge-to-judge reliability test.

Timing variables were measured with a custom Python timing tool while using the public web interfaces.\footnote{The timing tool is part of the repository's Data Collection Program. It records user-visible interaction phases from the outside of the public interface and does not expose provider-internal compute allocation or backend reasoning traces. Because the study used public web interfaces rather than APIs, timing was manually started and stopped from visible UI cues, such as prompt submission, the end of the visible reasoning phase, and completion of the final answer. An API-based or fully instrumented experiment would likely produce different timing measurements.} \texttt{LLM\ Response\ Time\ (s)} is the total time from submitting the prompt until the model finished writing its answer. \texttt{LLM\ Reasoning\ Time\ (s)} is the time from submitting the prompt until the model stopped showing its reasoning/thinking phase and began writing the final answer. Because all tested models were reasoning models, the reasoning timer starts at prompt submission and stops when the final answer begins. These timings describe the \textbf{user-visible interaction, not provider-internal compute time.}

\textbf{Social posting.} The project also adds a \textbf{public-distribution layer}. For each experiment, the four recordings are arranged into a unified horizontal \texttt{16:9} \texttt{2\ x\ 2} matrix comparison video. On X (Twitter), this horizontal composite was posted directly. On TikTok and YouTube Shorts, the same horizontal composite was fitted into the center of a \textbf{vertical video frame}, with unused space above and below the composite, and no additional platform-specific follow-up post. The \textbf{average composite-video duration} was \texttt{90.9} seconds across the \texttt{17} experiments. The matrix video position was fixed across experiments:

\[
V_e =
\begin{bmatrix}
v_{e,\text{GPT}} & v_{e,\text{Gemini}} \\
v_{e,\text{Grok}} & v_{e,\text{Claude}}
\end{bmatrix},
\]

where $v_{e,m}$ is the rendered recording for model family $m$ in experiment $e$. The four possible families are GPT, Gemini, Grok, and Claude. Each video quadrant $v_{e,m}$ was labeled with the specific model variant used in that run, such as GPT-5.2 Extended Thinking, Gemini 3 Pro, Grok 4.1 Thinking, or Opus 4.5 Thinking 32K, rather than only a broad family label. \textbf{Each post} uses a \textbf{matching AI-generated soundtrack} generated with the AI tool Suno,\footnote{Suno is the AI music-generation platform used to convert the generated song prompts and lyrics into the soundtrack paired with each comparison video. See https://suno.com/.} and most songs used the tracker style label \texttt{hip\ hop,\ pop}. The tracker also recorded social-packaging variables such as Suno version, song BPM, voice gender, song style, and video duration. The repository's Data Collection Program generated the song-lyrics prompt from the task prompt, human observations, and human scores; after that prompt was prepared, one of the \textbf{tracked lyric-generation LLMs produced the lyrics}.\footnote{The song-lyrics prompt template was generated by the repository's Data Collection Program from the experiment prompt, human observations, and human scores, then sent to the tracked lyric-generation LLM for that experiment. The template is reproduced in Appendix A.3.} The song prompt also required phonetic or nickname-style model names, such as "Gee-P-Tee," "Gem-in-eye," "Grok," or "Claw-d," so listeners on social media could understand the model references when hearing the soundtrack. This creates a shared social protocol so that viewers consume \textbf{standardized comparative content} rather than isolated clips.

\textbf{X (Twitter)} was the primary social-distribution channel and the \textbf{main reach outcome} modeled later in the paper, because it produced far \textbf{larger observed exposure} than TikTok or YouTube Shorts in this dataset. This platform difference should be interpreted alongside follower-base differences: the first experiment was posted with \texttt{3,570} X followers, \texttt{64} TikTok followers, and \texttt{955} YouTube followers, while the final experiment was posted with \texttt{5,143}, \texttt{95}, and \texttt{1,038} followers on those platforms, respectively. The larger observed X reach may therefore partly reflect the larger pre-existing X audience rather than platform mechanics alone. The \textbf{X posting format was also standardized}. The first post in each thread contained the experiment prompt text and the horizontal \texttt{16:9} composite matrix video $V_e$. A second post in the same thread provided the supporting experiment details: output line counts, model-specific scores, reasoning time, total response time, reasoning-word counts, the first-output-only/no-retry note, the repository link in github, custom-instruction notes, and timing caveats. That second post was accompanied by \textbf{four screenshots of the original public model interfaces}, one per model run, preserving visible evidence of the prompt, public chat UI, reasoning display, and generated output or code when visible. Those four interface screenshots are part of the public X thread evidence; the local experiment folders preserve the generated files, README metadata, and one preview image per experiment rather than four local public-interface screenshots.

\textbf{Data collection.} Social metrics were \textbf{recorded on a fixed 24-hour window after each post was published}. The main tracked outcomes were X impressions, likes, and shares, alongside TikTok views and likes and YouTube Shorts views and likes. Follower counts were also recorded at posting time so that early-account and later-account runs could be interpreted in context rather than treated as directly exchangeable exposure conditions.

The original task prompts, scoring prompt template, song prompt template, tracker, notebook, and figures are preserved in the repository and summarized again in the Appendix.

\section{Weighted Performance Model}\label{weighted-performance-model}

To compare outputs with a single scalar summary, I define the weighted performance score

\[
S_i = 0.40 \, PA_i + 0.35 \, FC_i + 0.25 \, UI_i.
\]

Neither the author nor Gemini directly assigned this weighted performance score. Both evaluators supplied the three component sub-scores (\texttt{PA}, \texttt{FC}, and \texttt{UI}), and the weighted performance score was then computed automatically in the tracker from those sub-scores. In the tracker, the human score column applies the Excel formula \texttt{ROUND(0.4*PA\ +\ 0.35*FC\ +\ 0.25*UI,\ 2)} to the human sub-score columns, while the Gemini weighted-performance column applies the same formula to Gemini's \texttt{PA}, \texttt{FC}, and \texttt{UI} sub-score columns. The human evaluator used a \texttt{0-10} scale with decimal values allowed, while Gemini was prompted for \texttt{1-10} integer scores; both sets of sub-scores were converted into tracker-computed weighted performance scores using the same formula.

These weights are \textbf{author-specified heuristic weights}, not values derived from formal theory or external validation. They were chosen to produce one practical summary score that prioritizes satisfying the prompt and making the application work, while still preserving UI quality as an important but secondary differentiator. Because that choice is subjective, the alternative-weight sensitivity analysis reported later should be treated as the primary robustness check.

\textbf{Reasoning efficiency} is then defined as performance earned per second of reasoning:

\[
E_i = \frac{S_i}{t_i^{\text{reason}}}.
\]

When reporting aggregate summaries, including experiment-level and family-level summaries, I average model-level efficiencies rather than dividing averaged performance by averaged reasoning time. This formulation makes the tradeoff between quality and latency explicit. The local sensitivity of the metric is

\[
\frac{\partial E_i}{\partial S_i} = \frac{1}{t_i^{\text{reason}}},
\qquad
\frac{\partial E_i}{\partial t_i^{\text{reason}}} =
- \frac{S_i}{\left(t_i^{\text{reason}}\right)^2}.
\]

The first derivative shows that performance gains matter more when reasoning time is short. The second shows that \textbf{efficiency drops nonlinearly as reasoning time increases}, with the largest marginal penalty at short reasoning times. In other words, low-latency successful outputs can produce very high efficiency, while added visible reasoning time must buy meaningful quality improvement to preserve the same efficiency.

For interpretive consistency across the analysis and discussion, I use author-chosen descriptive ranges rather than hard class labels or notebook-derived quartile thresholds:

\begin{itemize}
\tightlist
\item Performance Score: \texttt{Low\ <\ 6}, \texttt{Medium\ =\ 6\ to\ 7.99}, \texttt{High\ >=\ 8}.
\item Reasoning Efficiency: \texttt{Low\ <\ 0.40}, \texttt{Medium\ =\ 0.40\ to\ 0.79}, \texttt{High\ >=\ 0.80}.
\item X Impressions (experiment level): \texttt{Low\ =\ 214\ to\ 24,987}, \texttt{Medium\ =\ 24,988\ to\ 75,948}, \texttt{High\ =\ 75,949\ to\ 206,661}.
\item Lines of HTML: \texttt{Low\ =\ 108\ to\ 465}, \texttt{Medium\ =\ 466\ to\ 650.5}, \texttt{High\ >\ 650.5\ to\ 1,422}.
\end{itemize}

These ranges are descriptive and were used only to organize interpretation; they were not used as supervised labels in the predictive models. The performance-score and reasoning-efficiency bins are heuristic author choices, while the X-impressions and HTML-lines ranges summarize the observed dataset scale.

\section{Experiments and Analysis}\label{experiments-and-analysis}

This section reports the main results from the analysis and findings log.

\subsection{Model Performance and Compute Efficiency}\label{model-performance-and-compute-efficiency}

Under this fixed public-interface protocol, \textbf{Claude was the strongest and most stable family} overall. Across all \texttt{68} generations, \textbf{longer reasoning time did not predict better quality} in the aggregate. \textbf{Grok was the weakest and most volatile family}, and GPT and Gemini occupied a competitive middle tier while getting there in very different ways: \textbf{GPT tended to be slower and much more verbose}, while \textbf{Gemini was dramatically faster and more efficient.}

\textbf{Table 1. Model-family means for performance, verbosity, latency, and reasoning efficiency across all \texttt{68} outputs. Reasoning efficiency is averaged from per-generation efficiencies, not computed as mean performance divided by mean reasoning time.}

\begin{longtable}[]{@{}lrrrrr@{}}
\toprule\noalign{}
Model Family & Human Perf. /10 & HTML Lines & Reason Time (s) & Response Time (s) & Reason Eff. \\
\midrule\noalign{}
\endhead
\bottomrule\noalign{}
\endlastfoot
GPT & 7.96 & 714.94 & 80.71 & 247.18 & 0.31 \\
Gemini & 7.97 & 344.24 & 14.65 & 46.24 & 0.67 \\
Grok & 6.07 & 193.76 & 91.88 & 112.24 & 0.42 \\
Claude & 8.51 & 732.65 & 48.47 & 148.35 & 0.38 \\
\end{longtable}

Several patterns stand out. First, \textbf{Claude led all three human sub-scores and the weighted overall performance score}, while also showing the lowest variability. Second, \textbf{GPT and Claude were the two verbose families}, both averaging \textbf{more than \texttt{700} HTML lines} per generation. \textbf{Gemini produced much shorter code on average}, and \textbf{Grok was the most compact family by far}. Third, \textbf{Gemini achieved nearly the same mean performance as GPT} while using \textbf{only \texttt{14.65} reasoning seconds on average} versus \texttt{80.71} for GPT. This produced the \textbf{highest mean reasoning efficiency} in Table 1 (\texttt{0.67}) for Gemini, while \textbf{GPT and Claude had lower Reasoning efficiency} despite strong raw scores because they spent substantially more visible reasoning time per generation.

This pooled summary is also supported by prompt-paired evidence rather than by means alone:

\textbf{Table 2. Prompt-level top-score counts by evaluator on weighted performance scores.}

\begin{longtable}[]{@{}lrr@{}}
\toprule\noalign{}
Family & Human score & Gemini judge score \\
\midrule\noalign{}
\endhead
\bottomrule\noalign{}
\endlastfoot
Claude & 10 & 10 \\
GPT & 6 & 9 \\
Gemini & 1 & 5 \\
Grok & 1 & 0 \\
\end{longtable}

\begin{figure}[H]
\centering
\includegraphics[width=0.82\linewidth]{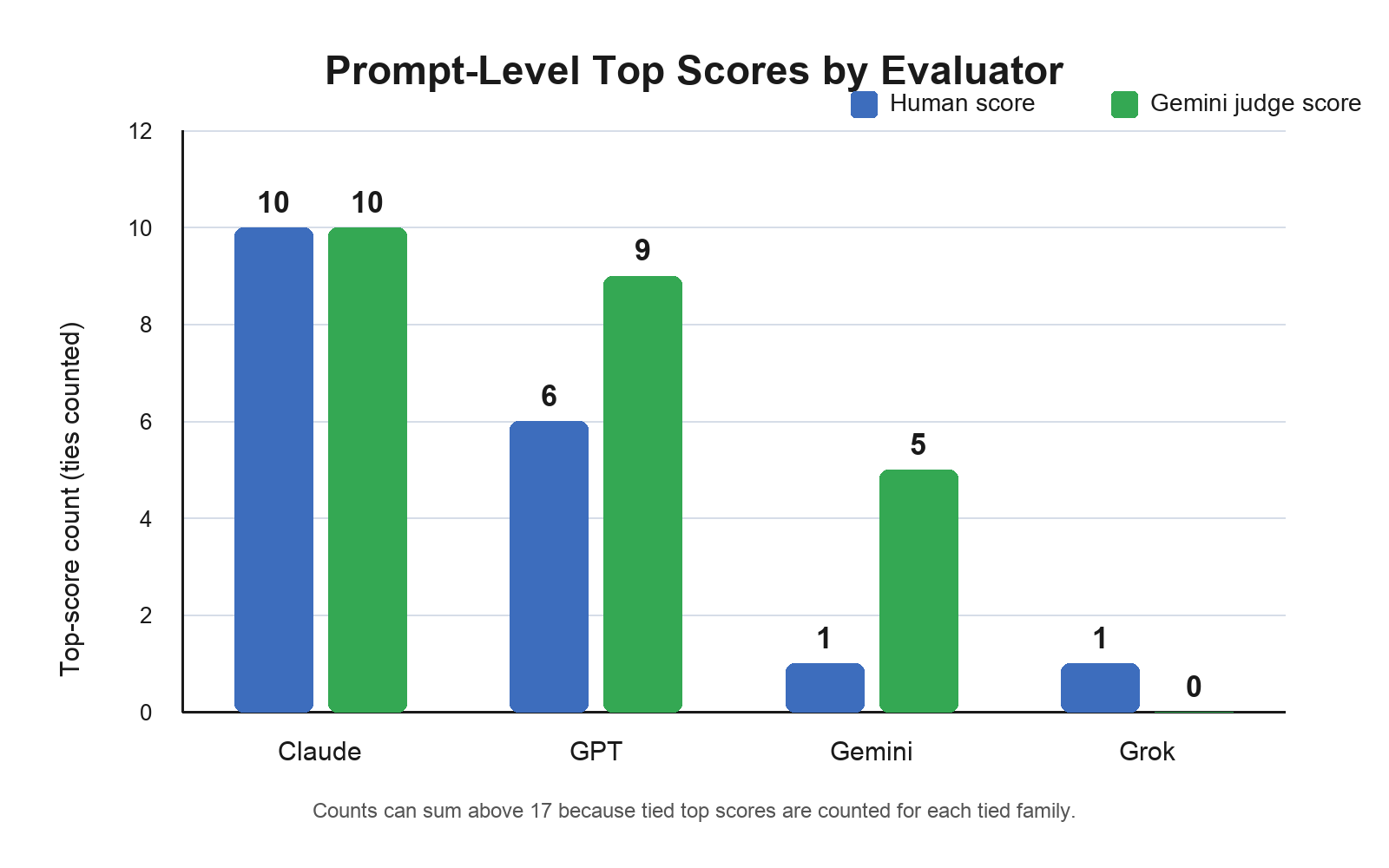}
\caption{Prompt-level top-score counts by evaluator on weighted performance scores. Counts can sum above 17 because tied top scores are counted for each tied family.}
\label{fig:figure-29}
\end{figure}

Table 2 and Figure 2 count tied top scores for every tied family. The human scores placed Claude top or tied for top in \texttt{10/17} prompts, including one tie, while Gemini's integer judge scores produced more tied top scores and placed Claude top or tied for top in \texttt{10} prompts, GPT in \texttt{9}, Gemini in \texttt{5}, and Grok in \texttt{0}.

\textbf{Table 3. Single-winner experiment counts on the primary human weighted performance score.}

\begin{longtable}[]{@{}lr@{}}
\toprule\noalign{}
Family & Experiment wins out of 17 \\
\midrule\noalign{}
\endhead
\bottomrule\noalign{}
\endlastfoot
Claude & 9 \\
GPT & 6 \\
Gemini & 1 \\
Grok & 1 \\
\end{longtable}

Table 3 gives the simpler single-winner summary across the \texttt{17} experiments using the primary human weighted performance score; the one tied human top score is assigned using the same single-winner convention as the notebook so the counts sum to \texttt{17}. In the paired human-score comparison, Claude beat GPT on \texttt{10/17} prompts (mean paired gap \texttt{+0.55}), Gemini on \texttt{12/17} prompts with \texttt{1} tie (mean paired gap \texttt{+0.54}), and Grok on \texttt{15/17} prompts (mean paired gap \texttt{+2.44}, Wilcoxon \texttt{p\ =\ 0.003}). The GPT and Gemini gaps are directionally favorable to Claude but not decisive in this small sample. Because the base score weights are heuristic, the sensitivity check using equal, adherence-heavy, function-heavy, and UI-heavy alternatives should be treated as the main robustness test. Under every scheme, \textbf{Claude still remained the top mean family}, with \texttt{8-9} prompt wins depending on the weighting. These checks strengthen the claim while keeping it narrow: Claude was strongest in this observational protocol, not necessarily under every possible controlled benchmark, while closer \textbf{GPT-versus-Gemini comparisons remain more weight-sensitive.}

\textbf{Longer measured reasoning time did not reliably translate into better performance scores.} Across the full dataset, the correlation between reasoning time and human performance was weak and non-significant (\texttt{Spearman\ rho\ =\ -0.094}, \texttt{p\ =\ 0.443}). A similar pattern held for Gemini's own overall scores. The slowest quartile of generations actually performed worse on average than the faster groups, which argues against a simple "more thinking means better code" story.

The clearest model-specific exception was Gemini itself. Within the \textbf{Gemini family}, \textbf{longer response times} were associated with \textbf{lower human-rated performance} (\texttt{Spearman\ rho\ =\ -0.577}, \texttt{p\ =\ 0.015}), and \textbf{longer reasoning times} were also associated with \textbf{lower human-rated performance} (\texttt{Spearman\ rho\ =\ -0.540}, \texttt{p\ =\ 0.025}). In this dataset, \textbf{Gemini's extra latency looked more like a struggle signal than a quality signal.}

Reasoning also consumed different fractions of total latency across families. Overall, \textbf{reasoning accounted for \texttt{40.4\%} of response time}, but \textbf{Grok spent \texttt{73.4\%} of its time in reasoning mode} on average. That did not translate into stronger outputs. Grok's mean efficiency was inflated by a few extremely short successful runs; its median reasoning efficiency was only \texttt{0.074}, the lowest family median in the study.

Finally, \textbf{code length mattered}, but only up to a point. \textbf{HTML length had only a weak positive relationship with human performance} (\texttt{Spearman\ rho\ =\ 0.243}, \texttt{p\ =\ 0.0456}). The practical pattern was that very short implementations underperformed, but \textbf{once a minimum level of structural completeness was reached}, simply \textbf{adding more lines did not guarantee better quality.}

\subsection{Latent Bias in LLM-as-a-Judge}\label{latent-bias-in-llm-as-a-judge}

The main concern here is \textbf{leniency rather than clearly proven self-preference}. The secondary Gemini judge was useful, but it was not neutral in the same way as the human score, and across the full dataset it \textbf{scored outputs higher than the human evaluator} on average, especially for \textbf{functional correctness and overall performance}.

\textbf{Table 4. Human-vs-Gemini judge score comparison across the full dataset.}

\begin{longtable}[]{@{}lrrrr@{}}
\toprule\noalign{}
Metric & Human Mean & Gemini Mean & Mean Gap & Wilcoxon \texttt{p} \\
\midrule\noalign{}
\endhead
\bottomrule\noalign{}
\endlastfoot
Prompt Adherence & 7.78 & 8.31 & +0.53 & 0.0678 \\
Functional Correctness & 7.58 & 8.37 & +0.78 & 0.0016 \\
UI Quality & 7.43 & 7.72 & +0.29 & 0.3426 \\
Overall Performance & 7.63 & 8.18 & +0.56 & 0.0162 \\
\end{longtable}

The \texttt{p} values in Table 4 are two-sided Wilcoxon signed-rank tests over paired Gemini-minus-human score differences across model outputs, with zero differences omitted from the signed-rank statistic.

The strongest and most reliable gap appeared in functional correctness. \textbf{Gemini tended to reward} technically \textbf{complete-looking behavior} even when the resulting application was \textbf{awkward, partially wrong, or weaker than the intended creative target}. This matters for browser applications because "it runs" is not the same as "it satisfies the prompt well."

I also explicitly investigated \textbf{whether Gemini favored its own outputs}. The results were more nuanced than the early draft of this paper assumed, and they should be interpreted cautiously because the paired self-bias analysis only has \texttt{17} experiment-level comparisons. Here, "overall performance" refers to the tracker-computed weighted performance score; \textbf{neither Gemini nor the human evaluator directly assigned that final scalar score}, since both supplied only the component sub-scores and the tracker computed the weighted result. It is also important to remember that \textbf{Gemini as a judge did not know which model generated each video}. In overall performance, the data did \textbf{not show strong evidence of a stable self-favoring pattern}, but that non-result should not be read as evidence that such bias is absent. \textbf{When Gemini rated Gemini outputs higher} on overall performance, the human evaluator agreed in all \texttt{11} such experiments. The strongest apparent self-boost appeared in \textbf{prompt adherence $PA$}, where \textbf{Gemini's own outputs received a larger mean internal advantage than the human gap suggested}, but the sample was too small to treat the non-significant test as resolving the question. UI was the least self-consistent area: when \textbf{Gemini favored itself on UI}, the \textbf{human evaluator disagreed about a quarter of the time.}

The main lesson is therefore not "Gemini always cheats for Gemini." The stronger and more defensible claim is that \textbf{Gemini was a systematically more lenient judge than the human evaluator.} The self-favoring question remains unresolved in this sample, so the practical benchmarking risk is clear leniency plus an underpowered self-bias test, not a clean proof either way.

\subsection{Social Reach and Predictive Modeling}\label{social-reach-and-predictive-modeling}

This section uses the tracker for \textbf{two exploratory supervised-learning questions}. The first is an \textbf{experiment-level} $e$ reach task: \textbf{given only information available before the 24-hour social media outcome window}, can a small \textbf{regression model predict 24-hour X impressions} for each social post? The second is a \textbf{generation-level} $i$ verbosity task: \textbf{given the prompt and model identity}, can a model \textbf{predict how many lines of HTML a single model output will contain?} These are not deployed forecasting systems; they are diagnostic models used to test whether the tracked variables contain predictive signal. For the \textbf{X-impressions} task, I used a \textbf{full-sample Lasso} screen followed by a \textbf{Ridge regression} evaluated with \textbf{leave-one-out cross-validation} on the \texttt{17} experiments. For the \textbf{HTML-lines task}, I compared \textbf{Ridge pipelines} under \textbf{leave-one-experiment-out validation} across the \texttt{68} model-level generations.

Two findings stand out here: the \textbf{tracked technical metrics did not predict 24-hour X reach} well in this dataset, while \textbf{HTML verbosity was predicted much more by model family than by prompt wording}.

\textbf{Social reach.} Social exposure was \textbf{highly concentrated and platform-dependent}. At the experiment level, \textbf{X (Twitter) averaged \texttt{50,188} impressions} in the first 24 hours, compared with \textbf{\texttt{408} TikTok views} and \textbf{\texttt{122} YouTube Shorts views}. On X, \textbf{\texttt{12} of the \texttt{17} experiments fell below \texttt{50,000} impressions}, while only \textbf{\texttt{5} exceeded that threshold}, revealing a strongly \textbf{right-skewed reach distribution}.

The \textbf{strongest pre-publication signal for X (Twitter)} outcomes was not technical quality but \textbf{baseline account state}. X follower count was strongly negatively correlated with both impressions and likes (\texttt{Pearson\ r\ =\ -0.828} for impressions), meaning earlier posts, when the account was smaller, often achieved the best reach-per-follower. \textbf{Variables creators often care about}, such as song BPM, voice gender, day of week, and posting time, did not show stable explanatory power in this sample. Because follower counts mostly reflected time and distribution context in this short longitudinal window, I treated them as descriptive context rather than retaining them in the restricted final technical/audio Ridge path.

\textbf{Predicting X impressions.} I trained an exploratory experiment-level $e$ \textbf{Ridge regression} model on a \textbf{log-transformed public-reach target}:

\[
y_e = \log(1 + x_e),
\]

where $x_e$ is the raw 24-hour X impressions for experiment $e$. Because the sample is only \texttt{17} experiments, this path should be read as exploratory and statistically underpowered rather than definitive. Leave-One-Out estimates on \texttt{17} points are inherently unstable, so the reported errors are more useful as descriptive evidence of weak predictive signal than as precise generalizable performance estimates. A full-sample Lasso screen [7] first narrowed a broader pre-24h pool, including distribution context, evaluation scores, and generation metrics. The final Ridge model [6] was then evaluated with Leave-One-Out Cross-Validation that refit scaling and Ridge inside each training fold after the full-sample exploratory feature screen, using only four pre-publication technical/audio variables. Because the feature screen was run on the full sample before this final model, the errors should be read as post-screen descriptive cross-validation estimates rather than nested feature-selection estimates:

\begin{itemize}
\tightlist
\item \texttt{ReasonTime\_mean}: the average visible public-interface reasoning/thinking time in seconds.
\item \texttt{Reasoning\_Ratio\_mean}: the average share of total response time spent thinking.
\item \texttt{Suno\_version}: binary encoding of the Suno version used for the soundtrack (\texttt{v5} vs. \texttt{v4.5}).
\item \texttt{Song\_BPM}: the tempo of the background AI-generated track.
\end{itemize}

For centered and scaled features, and omitting the intercept term for notational simplicity, the ridge estimator is

\[
\hat{\beta} =
\arg\min_{\beta} \|y - X \beta\|^2 + \alpha \|\beta\|^2,
\qquad
\hat{\beta} = (X^\top X + \alpha I)^{-1} X^\top y.
\]

Out of sample, the model remained weak:

\begin{itemize}
\tightlist
\item \texttt{MAE\ =\ 46,874} impressions
\item \texttt{RMSE\ =\ 72,633} impressions
\item \texttt{R\^{}2\ =\ -0.377}
\end{itemize}

The largest misses came from the biggest reach outliers, especially \texttt{drifting\ car}, \texttt{flappy\ bird}, \texttt{spaceship}, and \texttt{paper\ airplane} experiments. All four final coefficients were negative, with reasoning time as the strongest effect. In practical terms, the \textbf{model learned that slower and more reasoning-heavy experiments tended to underperform}, but it still \textbf{could not explain the main drivers of 24-hour X (Twitter) reach} in this dataset. This predictive path is therefore more valuable as descriptive negative evidence for this feature set and protocol than as a useful forecaster.

\textbf{Predicting HTML lines from prompt and model identity.} The second predictive task asked whether output length could be predicted from prompt and model identity. This supervised regression task used the \texttt{68} model-level generations $i$ and compared four Ridge pipelines under leave-one-experiment-out validation grouped by \texttt{Experiment\ ID}, so that the held-out rows always came from prompts unseen during training. The four pipelines were input words only, model family only, input words plus model family, and TF-IDF prompt text plus input words plus model family [6, 8, 9]. Prompt text was encoded with TF-IDF only in the prompt-aware comparison pipeline.

\textbf{Table 5. HTML-lines prediction under leave-one-experiment-out grouped validation.}

\begin{longtable}[]{@{}lrrr@{}}
\toprule\noalign{}
Model & MAE & RMSE & \texttt{R\^{}2} \\
\midrule\noalign{}
\endhead
\bottomrule\noalign{}
\endlastfoot
Model family only & 135.2 & 192.6 & 0.576 \\
Input words + model family & 139.6 & 208.8 & 0.501 \\
Prompt text + input words + model family & 148.5 & 215.0 & 0.471 \\
Input words only & 241.0 & 310.1 & -0.101 \\
\end{longtable}

\begin{figure}[H]
\centering
\includegraphics[width=0.95\linewidth]{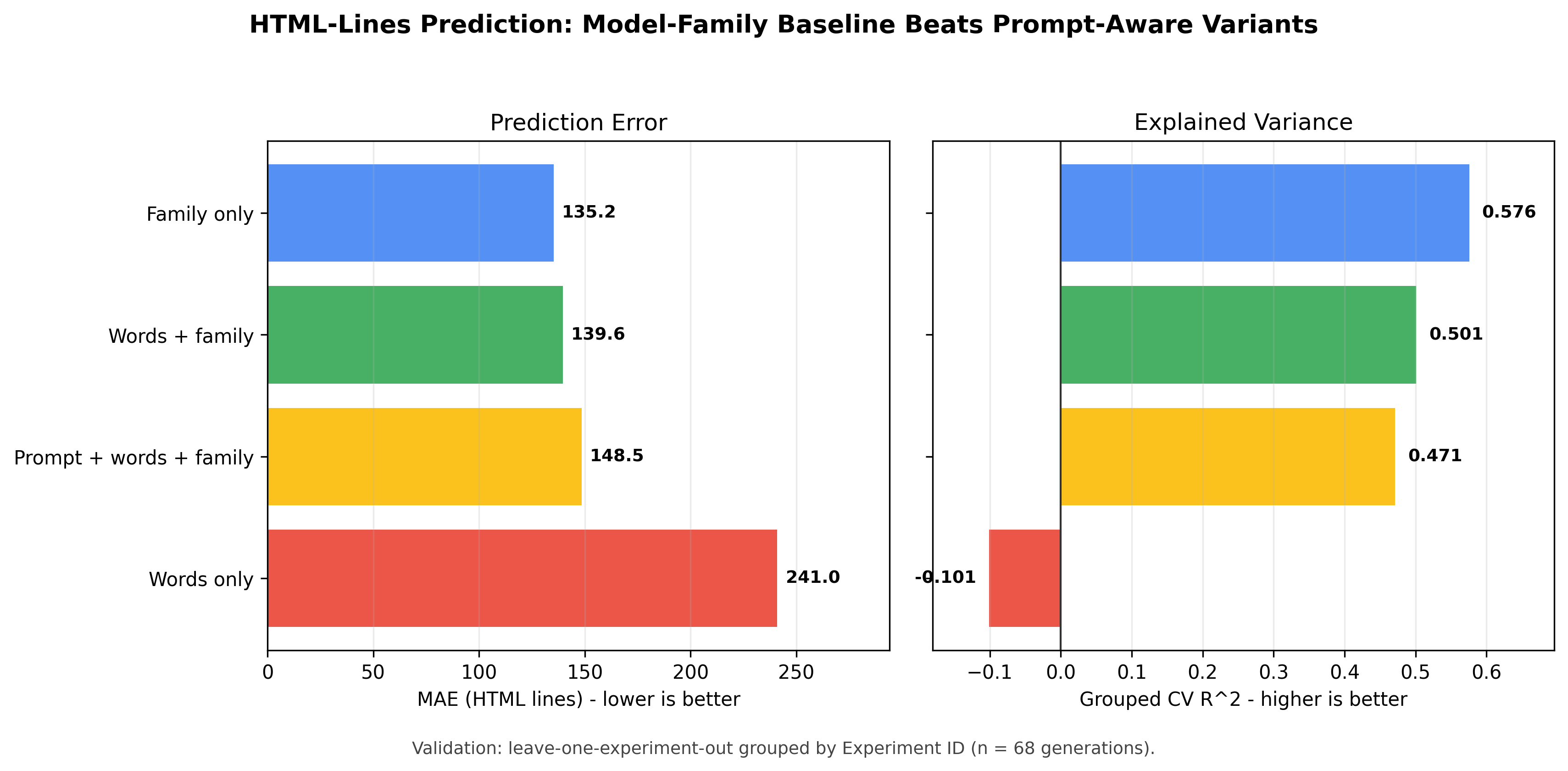}
\caption{HTML-lines model comparison under leave-one-experiment-out evaluation. The model-family-only baseline achieved the best out-of-sample tradeoff, outperforming the prompt-aware alternatives on both MAE and R\^{}2.}
\label{fig:figure-25}
\end{figure}

This was one of the \textbf{clearest results} in the study. \textbf{Model family} was substantially more \textbf{predictive of HTML verbosity than prompt wording}. Even adding input words and prompt text through TF-IDF features did not beat the simple family-only baseline. Descriptively, the same pattern was already visible before training: mean HTML length by family was \texttt{732.65} for Claude, \texttt{714.94} for GPT, \texttt{344.24} for Gemini, and \texttt{193.76} for Grok.

This second result matters because it contrasts sharply with the impressions task. In the impressions task, the \textbf{tracked technical metrics did not explain 24-hour X reach} well in this dataset. In the HTML-lines task, however, \textbf{model identity explained a large and stable part of the variance}, while input words and prompt text did not improve over the family-only baseline in grouped validation. Put differently, \textbf{prompt length and prompt text were weak levers for code verbosity} compared with the persistent \textbf{behavioral signature of the model family} itself.

The repository also includes \textbf{interactive Gradio playgrounds} for both trained paths, allowing exploratory use of the X-impressions and HTML-lines predictors without changing the underlying paper results.

\section{Conclusion}\label{conclusion}

This study shows that \textbf{first-output-only} evaluation reveals a very different picture of LLM web generation than prompt-engineering demos or iterative coding workflows. Under a strict one-shot public-interface protocol, \textbf{model family mattered a great deal}. Claude was the strongest and most consistent family overall, GPT and Gemini formed a competitive but behaviorally distinct middle tier, and \textbf{Grok was the most volatile and lowest-performing family} in the dataset. The \textbf{Claude lead was visible} not only in pooled means, but also in prompt-level wins and simple score-weight sensitivity checks.

\textbf{Longer reasoning did not improve quality overall.} In the Gemini family, longer reasoning and longer total response times were actually associated with worse human-rated performance. That makes \textbf{reasoning time a poor proxy for competence} in this setting. Similarly, \textbf{longer code was not a reliable route to better applications} once a minimum threshold of structural completeness had been reached.

The paper also addresses two predictive questions. First, the \textbf{tracked technical generation metrics were not sufficient to support a useful X-impressions predictor} in this small dataset; \textbf{factors outside the selected pre-publication technical and audio variables likely mattered}, including platform context and audience state. Second, \textbf{HTML output length was much more predictable}, and it was predicted mainly by model family; adding input words and prompt text did not improve over the family-only baseline in grouped validation. In other words, this dataset supports \textbf{prediction of HTML lines mostly from model identity}, but it does not support confident prediction of 24-hour X impressions from the tracked technical metrics alone.

Finally, the judge-comparison analysis suggests that \textbf{Gemini as a judge should be used carefully}. It was more lenient than the human evaluator, especially on functional correctness, but the evidence for a stable \textbf{self-favoring bias remained inconclusive} in this small sample. The more robust conclusion is that \textbf{human evaluation remains important when the target artifact is a visual}, interactive application rather than a narrow text or code completion task.

The main \textbf{limitations are the small experiment-level sample for social prediction,} public-interface version drift over time, public UI differences across providers, the use of LMArena for Claude, collection-time confounds, and the \textbf{reliance on one primary human scorer}. The study is therefore best interpreted as a \textbf{longitudinal observational analysis} rather than a definitive causal comparison of model capability. Even so, the longitudinal design, fixed user protocol, and fully reproducible notebook make the study a useful benchmark for future work on one-shot web generation. One-shot web generation matters because real users often need a complete, usable first answer, not a system that only looks strong after several rounds of repair.

\section{Ethics Statement}\label{ethics-statement}

This study was conducted independently and without external funding from any model provider. All model interactions were performed through public user-facing interfaces rather than private APIs or privileged access. The goal was to approximate real end-user behavior, not to benchmark internal capabilities under idealized lab conditions.

The social-media data used in the analysis consists only of public metrics from the author's own accounts, including impressions, likes, shares, and follower counts at posting time. No external private user data was collected. The generated applications, recordings, and posts were published transparently under the author's identity.

The paper also explicitly acknowledges the limitations of LLM-as-a-judge methods. Gemini was used as a secondary evaluator in isolated blind chats, and the study reports both its usefulness and its biases rather than treating model-judged scores as ground truth.

\section*{Reproducibility Statement}

All materials needed to reproduce the analysis are available in the repository [5]:

\begin{itemize}
\tightlist
\item \texttt{experiment\_tracker.xlsx}: the main structured dataset with \texttt{68} rows and \texttt{48} columns.
\item \texttt{experiments/}: per-experiment raw outputs, generated HTML files, README files, public post links where available, and one local preview image per experiment. The four public-interface screenshots used as run evidence were attached to the corresponding X threads rather than stored as four separate local files in each experiment folder.
\item \texttt{data\_collection\_program/}: the local tool used to standardize experiment setup, timing, output capture, scoring prompts, song-prompt templates, and posting-text preparation.
\item \texttt{research.ipynb}: the full notebook, including preprocessing, exploratory analysis, supervised modeling, and interactive prediction playgrounds.
\item \texttt{figures/} and \texttt{tables/}: exported visual assets used to support the manuscript.
\end{itemize}

The public social-distribution channels used in the study were the author's own accounts: X (Twitter), \href{https://x.com/diegocabezas01}{@diegocabezas01}; TikTok, \href{https://www.tiktok.com/@ai\_academy\_2030}{@ai\_academy\_2030}; and YouTube, \href{https://www.youtube.com/@airevolution23}{@airevolution23}. The X post links for individual experiments are preserved in the corresponding experiment-folder README files under the repository's \texttt{experiments/} directory.

The notebook reconstructs both data granularities used in the study:

\begin{itemize}
\tightlist
\item \texttt{df}: the model-level table with one row per generation.
\item \texttt{exp\_df}: the experiment-level table obtained by collapsing the four generations associated with each experiment.
\end{itemize}

The predictive setup is also fully specified:

\begin{itemize}
\tightlist
\item Exploratory X-impressions model: full-sample Lasso screened a broader pre-24h pool; the final post-screen LOOCV Ridge refit scaling and Ridge inside each training fold using \texttt{log(1\ +\ impressions)} and the restricted set \texttt{ReasonTime\_mean}, \texttt{Reasoning\_Ratio\_mean}, \texttt{Suno\_version}, and \texttt{Song\_BPM}.
\item HTML-lines model: grouped Ridge pipelines evaluated with leave-one-experiment-out validation comparing input words only, model family only, input words plus model family, and TF-IDF prompt text plus input words plus model family.
\end{itemize}

The weighted performance formula, reasoning efficiency equation, judge prompts, song prompt template, protocol figure, and prompt-paired robustness checks are all documented in this manuscript and in the repository. Running the notebook on the bundled tracker reproduces the analysis and the exported figures.

\section*{References}

[1] Chen, M., Tworek, J., Jun, H., et al. (2021). \emph{Evaluating Large Language Models Trained on Code}. arXiv:2107.03374.

[2] Zheng, L., Chiang, W.-L., Sheng, Y., et al. (2023). \emph{Judging LLM-as-a-Judge with MT-Bench and Chatbot Arena}. arXiv:2306.05685.

[3] Bubeck, S., Chandrasekaran, V., Eldan, R., et al. (2023). \emph{Sparks of Artificial General Intelligence: Early Experiments with GPT-4}. arXiv:2303.12712.

[4] Chang, Y., Wang, X., Wang, J., et al. (2024). \emph{A Survey on Evaluation of Large Language Models}. \emph{ACM Transactions on Intelligent Systems and Technology}, 15(3).

[5] Cabezas Palacios, D. (2026). \emph{HTML AI Battle: Dataset, Notebook, and Paper Materials}. GitHub repository: https://github.com/diegocp01/html\_ai\_battle

[6] Hoerl, A. E., and Kennard, R. W. (1970). \emph{Ridge Regression: Biased Estimation for Nonorthogonal Problems}. \emph{Technometrics}, 12(1), 55-67.

[7] Tibshirani, R. (1996). \emph{Regression Shrinkage and Selection via the Lasso}. \emph{Journal of the Royal Statistical Society: Series B}, 58(1), 267-288.

[8] Pedregosa, F., Varoquaux, G., Gramfort, A., et al. (2011). \emph{Scikit-learn: Machine Learning in Python}. \emph{Journal of Machine Learning Research}, 12, 2825-2830.

[9] Salton, G., and Buckley, C. (1988). \emph{Term-Weighting Approaches in Automatic Text Retrieval}. \emph{Information Processing and Management}, 24(5), 513-523.

[10] Zhou, S., Xu, F. F., Zhu, H., et al. (2023). \emph{WebArena: A Realistic Web Environment for Building Autonomous Agents}. arXiv:2307.13854.

[11] Koh, J. Y., Lo, R., Jang, L., et al. (2024). \emph{VisualWebArena: Evaluating Multimodal Agents on Realistic Visual Web Tasks}. arXiv:2401.13649.

\section*{Appendix}

\subsection*{A.1 Notebook Data-Exploration Question Summary}

The notebook's \texttt{1.3\ -\ Data\ Exploration\ (EDA)} section contains additional descriptive checks that informed the paper. The main text uses the strongest of these checks directly, especially the model-family summary, reasoning-time analysis, Gemini-judge analysis, X-reach correlations, and HTML-lines prediction setup. For completeness, the full question-level EDA summary is preserved here. These checks are descriptive and should be read as exploratory because many were run on only \texttt{68} model-level generations or \texttt{17} experiment-level posts.

\textbf{Model-level questions.}

\begin{itemize}
\tightlist
\item \textbf{Does code in reasoning correlate with a higher final output score?} Outputs with visible code in the reasoning area had higher mean human performance (\texttt{8.49} vs. \texttt{7.38}) and higher mean Gemini performance (\texttt{8.82} vs. \texttt{8.00}), but the human-score comparison was borderline rather than decisive (\texttt{Mann-Whitney\ p\ =\ 0.0554}). The pattern is also confounded by model family: \texttt{14/15} code-in-reasoning cases came from Claude.
\item \textbf{Are lines of HTML related to final human and Gemini performance?} HTML length had a weak positive relationship with human performance (\texttt{Spearman\ rho\ =\ 0.243}, \texttt{p\ =\ 0.0456}) and a stronger positive relationship with Gemini performance (\texttt{Spearman\ rho\ =\ 0.401}, \texttt{p\ =\ 0.0007}). The practical interpretation is not that more code is automatically better, but that extremely short outputs often lacked enough structure.
\item \textbf{What is the reasoning-efficiency metric?} Reasoning efficiency was computed as performance divided by reasoning seconds at the model-generation level. Across all \texttt{68} outputs, human-based reasoning efficiency had mean \texttt{0.444} and median \texttt{0.354}; Gemini-based reasoning efficiency had mean \texttt{0.477} and median \texttt{0.422}.
\item \textbf{Are longer prompts associated with higher performance, higher efficiency, or more HTML lines?} Longer prompts were not associated with better human or Gemini performance. They were associated with lower reasoning efficiency (\texttt{Spearman\ rho\ =\ -0.305}, \texttt{p\ =\ 0.0115} for human efficiency; \texttt{rho\ =\ -0.340}, \texttt{p\ =\ 0.0045} for Gemini efficiency). Prompt length had a weak Pearson association with HTML lines (\texttt{r\ =\ 0.240}, \texttt{p\ =\ 0.0483}), but the rank correlation was not stable (\texttt{rho\ =\ 0.075}, \texttt{p\ =\ 0.5462}).
\item \textbf{Are lines of HTML related to reasoning efficiency?} No meaningful relationship appeared. Human reasoning efficiency had \texttt{Spearman\ rho\ =\ -0.005} (\texttt{p\ =\ 0.9676}) with HTML length, and Gemini reasoning efficiency had \texttt{rho\ =\ 0.022} (\texttt{p\ =\ 0.8573}).
\item \textbf{Are experiments similar in HTML-verbosity patterns?} Yes. Across the four model-family HTML-length vectors, pairwise experiment cosine similarity was high on average (\texttt{mean\ =\ 0.975}, minimum \texttt{0.910}, maximum \texttt{0.999}). This reflects a consistent family-level verbosity pattern: Claude and GPT usually produced the largest share of code, while Grok was consistently the most compact.
\item \textbf{For the exact same prompt, how far apart were the shortest and longest implementations?} The within-experiment HTML spread was large. The average range between shortest and longest output was \texttt{616.59} lines, with a maximum spread of \texttt{1,064} lines. Grok was the shortest implementation in all \texttt{17/17} experiments, while the longest implementation was split between Claude (\texttt{9/17}) and GPT (\texttt{8/17}).
\item \textbf{What minimum amount of HTML lines appears to support good performance?} There was no hard guarantee. The shortest good output had only \texttt{108} lines and scored \texttt{9.50}, but the \texttt{<=200}-line bucket had only a \texttt{30.8\%} good-output rate. The strongest practical bucket was \texttt{401-500} lines, with an \texttt{85.7\%} good-output rate. A threshold of \texttt{261} lines was the smallest cutoff that reached a \texttt{70\%} good-output rate among outputs at or above the cutoff.
\item \textbf{Does prompt length correlate with model quality outcomes?} No clear positive quality relationship appeared. Prompt length had weak or negative rank correlations with the score dimensions, including human prompt adherence (\texttt{Spearman\ rho\ =\ -0.243}, \texttt{p\ =\ 0.0461}) and overall human performance (\texttt{rho\ =\ -0.054}, \texttt{p\ =\ 0.6623}).
\item \textbf{Does Gemini score itself better, and is that because Gemini outputs were better or because Gemini was biased?} The EDA did not show a clean, statistically decisive self-favoring result. Gemini's judge score favored Gemini's own output on overall performance in \texttt{11/17} experiments, and in those cases the human evaluator also preferred Gemini \texttt{100\%} of the time. The most suspicious self-boost was in prompt adherence, where Gemini's own-output advantage was larger than the human advantage, but the bias test was not significant (\texttt{p\ =\ 0.6050}).
\item \textbf{What percentage of response time was reasoning?} Reasoning accounted for \texttt{40.36\%} of response time overall. Grok was the outlier, spending \texttt{73.37\%} of total response time in reasoning on average, compared with \texttt{31.21\%} for Claude, \texttt{25.55\%} for GPT, and \texttt{31.31\%} for Gemini.
\item \textbf{Is reasoning time related to performance?} Overall, no. Reasoning time had weak negative, non-significant relationships with human performance (\texttt{Spearman\ rho\ =\ -0.0945}, \texttt{p\ =\ 0.4434}) and Gemini performance (\texttt{rho\ =\ -0.1497}, \texttt{p\ =\ 0.2230}). The slowest reasoning-time quartile had the lowest mean human performance (\texttt{6.58}).
\item \textbf{Is total inference time related to performance?} Overall, no. Total response time had near-zero relationships with human performance (\texttt{Spearman\ rho\ =\ 0.056}, \texttt{p\ =\ 0.6502}) and Gemini performance (\texttt{rho\ =\ 0.067}, \texttt{p\ =\ 0.5851}). Within Gemini, longer total response time was associated with lower human performance (\texttt{Spearman\ rho\ =\ -0.577}, \texttt{p\ =\ 0.0153}).
\item \textbf{What does the per-model summary show?} Claude led the mean overall human score (\texttt{8.51}) and had the lowest overall-score standard deviation (\texttt{1.62}). GPT (\texttt{7.96}) and Gemini (\texttt{7.97}) were close in mean performance but differed sharply in latency and verbosity. Grok had the lowest mean score (\texttt{6.07}) and highest volatility (\texttt{std\ =\ 3.28}).
\item \textbf{Which experiment took each model the longest reasoning and inference time?} The longest GPT reasoning and total-response run was \texttt{backflip\_1014} (\texttt{466} seconds reasoning, \texttt{808} seconds response). Grok's longest reasoning and response run was \texttt{basketball\_0131} (\texttt{232} seconds reasoning, \texttt{252} seconds response). Claude's longest reasoning and response run was \texttt{backflip\_1014} (\texttt{162} seconds reasoning, \texttt{265} seconds response). Gemini's longest reasoning run was \texttt{basketball\_0131} (\texttt{24} seconds), while its longest total response was \texttt{snake\_0121} (\texttt{69} seconds).
\item \textbf{Which experiment produced each model family's longest HTML output?} GPT's longest output was \texttt{snake\_0121} (\texttt{1,422} lines), Claude's was \texttt{airplane\_0117} (\texttt{1,254} lines), Gemini's was \texttt{paper\_airplane\_0204} (\texttt{509} lines), and Grok's was \texttt{snake\_0121} (\texttt{366} lines).
\item \textbf{What is the direct correlation of code length versus human performance?} The repeated code-length check matched the broader result: Pearson \texttt{r\ =\ 0.185} (\texttt{p\ =\ 0.1309}) and Spearman \texttt{rho\ =\ 0.243} (\texttt{p\ =\ 0.0456}). The relationship is weak and should be interpreted as a minimum-completeness signal rather than a monotonic "more code is better" law.
\end{itemize}

\textbf{Experiment-level questions.}

\begin{itemize}
\tightlist
\item \textbf{Are longer prompts associated with weaker or inconsistent gains in X impressions?} Prompt length was not meaningfully associated with X impressions (\texttt{Pearson\ r\ =\ -0.269}, \texttt{p\ =\ 0.2961}; \texttt{Spearman\ rho\ =\ -0.274}, \texttt{p\ =\ 0.2864}). The longest-prompt quartile had lower mean impressions than the shortest and third quartiles, but the pattern was unstable in \texttt{17} experiments.
\item \textbf{How did X follower count relate to impressions and likes?} Starting X follower count was strongly negatively correlated with raw impressions (\texttt{Pearson\ r\ =\ -0.828}, \texttt{p\ <\ 0.0001}) and likes (\texttt{r\ =\ -0.720}, \texttt{p\ =\ 0.0011}). Earlier posts from the smaller account period had the highest reach per follower, so follower count is better read as a time/context marker than a simple audience-size advantage.
\item \textbf{Is song BPM related to impressions?} No stable relationship appeared (\texttt{Pearson\ r\ =\ -0.257}, \texttt{p\ =\ 0.3195}; \texttt{Spearman\ rho\ =\ -0.163}, \texttt{p\ =\ 0.5325}).
\item \textbf{Is model-level generation time related to performance?} Generation hour varied at the model-row level, so the notebook analyzed all \texttt{68} generations rather than using an arbitrary first model per experiment. The relationship was weak or absent for human performance (\texttt{Pearson\ r\ =\ -0.168}, \texttt{p\ =\ 0.1719}; \texttt{Spearman\ rho\ =\ 0.032}, \texttt{p\ =\ 0.7984}) and Gemini performance (\texttt{Pearson\ r\ =\ 0.028}, \texttt{p\ =\ 0.8241}; \texttt{Spearman\ rho\ =\ 0.052}, \texttt{p\ =\ 0.6747}).
\item \textbf{Does song voice gender influence X impressions?} No significant difference appeared. Male-voice songs had mean \texttt{48,420} impressions and female-voice songs had mean \texttt{52,713}; the Mann-Whitney test was not significant (\texttt{p\ =\ 0.6961}).
\item \textbf{Are day of week and posting time on X correlated with impressions?} The observed Wednesday/Saturday difference was not significant (\texttt{Mann-Whitney\ p\ =\ 0.7001}). Posting hour also did not show a significant relationship with impressions (\texttt{Pearson\ r\ =\ -0.169}, \texttt{p\ =\ 0.5156}; \texttt{Spearman\ rho\ =\ -0.432}, \texttt{p\ =\ 0.0834}).
\item \textbf{Is song style correlated with X impressions?} The notebook found a significant Mann-Whitney result (\texttt{p\ =\ 0.0440}), but it is not robust because \texttt{15/17} experiments used \texttt{hip\ hop,\ pop} and only \texttt{2/17} used the longer trap/rap style label. The small group size and time-dependent reach changes make this result unstable.
\item \textbf{Does prompt length correlate with X impressions?} This repeated the experiment-level prompt-length check and found no meaningful relationship (\texttt{Pearson\ r\ =\ -0.269}, \texttt{p\ =\ 0.2961}; \texttt{Spearman\ rho\ =\ -0.274}, \texttt{p\ =\ 0.2864}).
\item \textbf{Are visual type and experiment type correlated with X impressions?} Neither relationship was significant. Simple visual tasks averaged \texttt{57,510} impressions versus \texttt{26,392} for complex tasks, but the point-biserial test was not significant (\texttt{p\ =\ 0.4112}). Game tasks averaged \texttt{53,498} impressions versus \texttt{49,170} for simulations, with no meaningful difference (\texttt{p\ =\ 0.9100}).
\item \textbf{Which variables had the largest linear correlations with X impressions?} As expected, X likes (\texttt{r\ =\ 0.955}) and X shares (\texttt{r\ =\ 0.909}) moved with X impressions because they are same-platform engagement outcomes. Among pre-publication or context variables, the strongest correlations were negative follower/context variables: X followers (\texttt{r\ =\ -0.828}), TikTok followers (\texttt{r\ =\ -0.798}), YouTube followers (\texttt{r\ =\ -0.797}), response-time mean (\texttt{r\ =\ -0.489}), and reasoning-time mean (\texttt{r\ =\ -0.457}). These correlations motivated the later caution that the X-impressions model is descriptive and underpowered.
\end{itemize}

\subsection*{A.2 Per-Experiment Task Prompts}

Each generation experiment used one task-specific natural-language prompt requesting a complete single-file HTML/CSS/JavaScript application. The 17 prompts used in the study are reproduced below in experiment order.

\begin{itemize}
\tightlist
\item \textbf{\texttt{flappy\_bird\_1210}}: Make a game like flappy bird using HTML/CSS/JS in a single HTML file
\item \textbf{\texttt{drifting\_car\_1213}}: Create a 3D simulation of a formula 1 car performing a continuous drifting donut in the street using HTML/CSS/JS in a single HTML file.
\item \textbf{\texttt{blooming\_flower\_1217}}: Create an animation of a single flower blooming in a pot. Using HTML/CSS/JS in a single HTML file.
\item \textbf{\texttt{dvd\_logo\_1220}}: Create a physics-accurate “Bouncing DVD Logo” simulation using HTML/CSS/JS in a single HTML file. The logo must bounce perfectly off all edges and corners, change color on every collision, and adapt correctly to dynamic window resizing.
\item \textbf{\texttt{super\_mario\_1224}}: Make a side-scrolling platformer game like Super Mario Bros. using HTML/CSS/JS in a single HTML file.
\item \textbf{\texttt{spaceship\_1227}}: Create a rocket launch animation starting with engine ignition and heavy smoke, followed by a slow liftoff that accelerates, using HTML/CSS/JS in a single HTML file.
\item \textbf{\texttt{new\_years\_1231}}: Create an interactive New Year's Eve fireworks display using HTML/CSS/JS in a single HTML file. When the user clicks, a rocket should launch and explode into colorful particles with gravity and drag physics. Include a 'Grand Finale' button that automatically launches a barrage of fireworks that form the glowing text 'HELLO 2026' in the sky.
\item \textbf{\texttt{dog\_fetch\_0103}}: Animate a dog fetching a ball. Using HTML/CSS/JS in a single HTML file.
\item \textbf{\texttt{gravity\_cloth\_0107}}: Create an interactive 'Tearable Cloth' simulation using HTML/CSS/JS in a single file. Render a grid of points connected by constraints (springs) that react to gravity. Allow the user to drag the cloth with their mouse and right-click (or hold Shift) to slice/cut the connections.
\item \textbf{\texttt{pirate\_ship\_0110}}: Create a simulation of a pirate ship sailing through open ocean. Using HTML/CSS/JS in a single HTML file.
\item \textbf{\texttt{backflip\_1014}}: Create a slow-motion simulation of a stick-figure human performing a standing backflip using HTML/CSS/JS. The figure needs articulated joints (knees, hips, elbows). It should crouch, jump, tuck tight to spin faster (conservation of angular momentum), and then extend legs to stick the landing.
\item \textbf{\texttt{airplane\_0117}}: Create a 3D simulation of an airplane landing using HTML/CSS/JS.
\item \textbf{\texttt{snake\_0121}}: Create a classic "Snake" game using HTML, CSS, and JavaScript, all within a single HTML file. The player controls a snake that grows longer when it eats food. The game ends if the snake collides with the walls or its own tail. Prioritize a rich, visually detailed UI, avoid simplistic visuals. Instead, aim for a realistic, pseudo-3D aesthetic for the snake, food, and background, using advanced visual effects, shadows, gradients, and animations to simulate depth and texture. Dont do simple black backgrounds or neon stuff, I want this too look like a CGI movie, the snake should looks like a real 3d depth snake, the food should look realistic. IMPORTANT: dont do a simple visual game, the realistic visuals are the main focus.
\item \textbf{\texttt{dna\_0124}}: Create a rotating 3D animation of a Double Helix Human DNA strand using HTML/CSS/JS in a single HTML file.
\item \textbf{\texttt{spongebob\_0128}}: Create an animation of SpongeBob and his home behind him, under the water using HTML/CSS/JS in a single HTML file.
\item \textbf{\texttt{basketball\_0131}}: Create a simulation of a basketball free-throw shot using HTML/CSS/JS in a single HTML file. When the user clicks, the ball should launch with a realistic parabolic arc, hit the backboard, and swish through the net.
\item \textbf{\texttt{paper\_airplane\_0204}}: Create a simulation of a paper airplane gliding through the air using HTML/CSS/JS in a single HTML file.
\end{itemize}

\subsection*{A.3 Song Prompt Template}

Each video is accompanied by an AI-generated song whose lyrics are generated by one of the tracked LLMs and then converted into audio with the selected Suno version. The lyric-generation model varied across experiments: Gemini 3 Pro, Opus 4.5 Thinking 32K, or GPT-5.2 Extended Thinking. The prompt template kept the same structure: provide the task prompt, observations, scores, and a reference style, then require phonetic or nickname-style model names for audio clarity. The exact prompt string used in practice is reproduced verbatim below. The model-name fields were filled with the current model labels for each run; the concrete template instance below shows the GPT label used when GPT-5.1 was current, while later GPT runs used the corresponding GPT-5.2 label.

\begin{Shaded}

\begin{Highlighting}

# Context

I have run an experiment comparing 4 LLMs (GPT, Gemini, Grok, and Claude Opus). They were all given the same prompt to generate HTML code. The prompt the models used today was ‘[prompt]’



# Task

Write a song summarizing the results of this week's experiment based on the provided "Observations" and "Scores."



# Critical Constraints

1. **Naming Convention:** Do NOT use the exact commercial names of the models. Instead, write their names **phonetically** or use nicknames that fit the flow and rhyme scheme of the song (e.g., "Gem-in-eye," "Gee-P-Tee," "Grok," "Claw-d").

2. **Content:** The lyrics must reflect the specific observations provided and the exact behaviour/output of the model (e.g., if one failed the HTML structure, mock it; if one was perfect, praise it).

3. **Structure:** Verse-Chorus structure.



# Input Data (Observations & Scores)

**Model 1: [Model name] GPT-5.1 Extended Thinking**

* Observations: [observations]

* Score: [score]/10



**Model 2: [Model name] Gemini 3 Pro**

* Observations: [observations]

* Score: [score]/10 



**Model 3: [Model name] Grok 4.1 Thinking**

* Observations: [observations]

* Score: [score]/10



**Model 4: [Model name] Claude Opus 4.5 Thinking (32k)**

* Observations: [observations]

* Score: [score]/10



# Reference Style

See an example of how the models sound good phonetically, in a previously generated song, and the style:

"[Verse 1] Gee-P-Tee Five point Two, Extended Thinking, in the frame, You just came out two days ago, but it’s a shame. You lacking my dog, my boi, what’s wrong with the head? There is no simulation, just parameters instead. You shoulda focused on the motion, not the text input, Score is zero Point three outta ten, yeah you stuck in the soot.



[Verse 2] Gem-in-eye Three Pro, nice intro right there, But what is wrong with the front tires? That’s a scare! Nice Tokyo Drift fam, you sliding with the pace, Score is Eight point eight outta ten, yeah you stayed in the race.



[Verse 3] G-rok Four point One Thinking, well at least you made it render, But the car is floating up like a Space-X sender. Instead of the car drifting, the camera moves away, Score is One outta ten, physics took a holiday.



[Verse 4] Claw-d O-pus Four point Five,  thinking Thirty two -K on the deck, I am completely speechless, gotta give respect. You didn’t just fulfill it, you made it interact, Playing in another dimension, that’s a fact. Compared to these little guys, homie, you the king, Beat the score board, you mastered everything. Score is 10 outta ten. ”

\end{Highlighting}

\end{Shaded}

\subsection*{A.4 Gemini Judging Prompt Template}

Each application video was scored in a separate Gemini chat using the video recording as input. Gemini did not see the other candidates in the comparison and was not told which model produced the clip being evaluated. The judging prompt content used in practice is reproduced below; the final scorecard layout is spaced as a readable three-column table.

\begin{Shaded}

\begin{Highlighting}

Role:



You are a Senior Frontend QA Engineer and Lead UI/UX Designer. Your task is to evaluate a video recording of a web application generated by an AI. You must be critical, precise, and objective.



Input Data:



The Goal: The original prompt used to generate the app: "[PROMPT used for the experiment]"



The Evidence: The attached video file showing the app in action.



Evaluation Criteria:



Score the application in the following three categories on a strict scale of 1-10 (integers only).



1. Prompt Adherence Score (PA)



Definition: Measures how completely the output implements the specified behavior and constraints found in "The Goal."



Scoring Guide:



10: Every requested feature (e.g., specific buttons, specific colors, specific mechanics) is present and accurate.



5: Main features are present, but specific constraints (like color or layout requests) were ignored.



1: The resulting app ignores the prompt entirely.



2. Functional Correctness Score (FC)



Definition: Evaluates the stability, physics, and logic of the application based on visual evidence.



Scoring Guide:



10: Physics mimic reality (if applicable), controls are responsive, logic flows correctly, and there are no visible glitches or "broken" states.



5: The app functions but feels "janky" or has minor logic errors (e.g., collision detection is loose).



1: The app crashes, freezes, or the core functionality is broken.



3. UI Quality Score (UI)



Definition: Assesses visual design, layout, readability, color hierarchy, and clarity of controls.



Scoring Guide:



10: Professional, modern, accessible design. Good use of whitespace, consistent color palette, and clear typography.



5: "Engineer Art." It works, but uses raw default HTML styles, clashing colors, or poor spacing.



1: Unusable layout, overlapping text, or painful color contrast.



Output Format:



Please provide your evaluation in the following structure:



Analysis



Final Scorecard



Metric | Score (1-10) | Reasoning



Prompt Adherence (PA) | [X] | [One sentence justification]



Functional Correctness (FC) | [X] | [One sentence justification]



UI Quality (UI) | [X] | [One sentence justification]

\end{Highlighting}

\end{Shaded}

\subsection*{A.5 Interpretive Ranges}

\begin{itemize}
\tightlist
\item Performance Score: \texttt{Low\ <\ 6}, \texttt{Medium\ =\ 6\ to\ 7.99}, \texttt{High\ >=\ 8}.
\item Reasoning Efficiency: \texttt{Low\ <\ 0.40}, \texttt{Medium\ =\ 0.40\ to\ 0.79}, \texttt{High\ >=\ 0.80}.
\item X Impressions: \texttt{Low\ =\ 214\ to\ 24,987}, \texttt{Medium\ =\ 24,988\ to\ 75,948}, \texttt{High\ =\ 75,949\ to\ 206,661}.
\item Lines of HTML: \texttt{Low\ =\ 108\ to\ 465}, \texttt{Medium\ =\ 466\ to\ 650.5}, \texttt{High\ >\ 650.5\ to\ 1,422}.
\end{itemize}

These are descriptive bins for reader interpretation, not data-derived class labels. The performance-score and reasoning-efficiency cutoffs are author-chosen heuristics rather than notebook-derived quartiles.

\subsection*{A.6 Repository Appendix Materials}

The exported visuals are stored in \texttt{figures/} and \texttt{tables/}. The interactive prediction playgrounds for the two fitted models are implemented inside Section 3 of \texttt{research.ipynb}.

\end{document}